\documentclass{article}

\usepackage{microtype}
\usepackage{graphicx}
\usepackage{subfigure}
\usepackage{booktabs} % for professional tables
\usepackage{diagbox}

\usepackage{hyperref}

\usepackage[accepted]{icml2025}

\usepackage{amsmath}
\usepackage{amssymb}
\usepackage{mathtools}
\usepackage{amsthm}
\usepackage{multirow}
\usepackage[capitalize,noabbrev]{cleveref}

\theoremstyle{plain}
\newtheorem{theorem}{Theorem}[section]

\newtheorem{corollary}[theorem]{Corollary}
\theoremstyle{definition}

\theoremstyle{remark}

\usepackage[textsize=tiny]{todonotes}

\icmltitlerunning{Submission and Formatting Instructions for ICML 2025}

\begin{document}

\twocolumn[
\icmltitle{DSSD: Efficient Edge-Device LLM Deployment and Collaborative Inference via Distributed Split Speculative Decoding}

% It is OKAY to include author information, even for blind
% submissions: the style file will automatically remove it for you
% unless you've provided the [accepted] option to the icml2025
% package.

% List of affiliations: The first argument should be a (short)
% identifier you will use later to specify author affiliations
% Academic affiliations should list Department, University, City, Region, Country
% Industry affiliations should list Company, City, Region, Country

% You can specify symbols, otherwise they are numbered in order.
% Ideally, you should not use this facility. Affiliations will be numbered
% in order of appearance and this is the preferred way.
\icmlsetsymbol{equal}{*}

\begin{icmlauthorlist}
\icmlauthor{Jiahong NING}{equal,yyy}
\icmlauthor{Ce ZHENG}{equal,comp}
\icmlauthor{Tingting YANG}{yyy,comp}
%\icmlauthor{}{sch}
%\icmlauthor{}{sch}
%\icmlauthor{}{sch}
\end{icmlauthorlist}

\icmlaffiliation{yyy}{Dalian Maritime University, Dalian, China}
\icmlaffiliation{comp}{Pengcheng Laboratory, Shenzhen, China}

% \icmlcorrespondingauthor{Ce ZHENG}{zhengc@pcl.ac.an}
\icmlcorrespondingauthor{Ce ZHENG}{zhengc@pcl.ac.cn}

% You may provide any keywords that you
% find helpful for describing your paper; these are used to populate
% the "keywords" metadata in the PDF but will not be shown in the document
\icmlkeywords{Machine Learning, ICML}

\vskip 0.3in
]

% this must go after the closing bracket ] following \twocolumn[ ...

% This command actually creates the footnote in the first column
% listing the affiliations and the copyright notice.
% The command takes one argument, which is text to display at the start of the footnote.
% The \icmlEqualContribution command is standard text for equal contribution.
% Remove it (just {}) if you do not need this facility.

%\printAffiliationsAndNotice{}  % leave blank if no need to mention equal contribution
\printAffiliationsAndNotice{\icmlEqualContribution} % otherwise use the standard text.

\begin{abstract}
Large language models (LLMs) have transformed natural language processing but face critical deployment challenges in device-edge systems due to resource limitations and communication overhead. To address these issues, collaborative frameworks have emerged that combine small language models (SLMs) on devices with LLMs at the edge, using speculative decoding (SD) to improve efficiency. However, existing solutions often trade inference accuracy for latency or suffer from high uplink transmission costs when verifying candidate tokens. In this paper, we propose Distributed Split Speculative Decoding (DSSD), a novel architecture that not only preserves the SLM–LLM split but also partitions the verification phase between the device and edge. In this way, DSSD replaces the uplink transmission of multiple vocabulary distributions with a single downlink transmission, significantly reducing communication latency while maintaining inference quality. Experiments show that our solution outperforms current methods, and codes are at: \href{https://github.com/JasonNing96/DSSD-Efficient-Edge-Computing}{https://github.com/JasonNing96/DSSD-Efficient-Edge-Computing}

\end{abstract}

\section{Introduction}

\textbf{Large language models (LLMs)} have revolutionized natural language processing, enabling powerful applications such as conversational agents, machine translation, and code generation \cite{chen2024personalizing}. 
Despite their capabilities, the LLMs face significant challenges across both devices and cloud. On devices, stringent constraints such as limited memory capacity, restricted battery, and computational power hinder the adoption of traditional LLM frameworks. Cloud deployments, while benefiting from scalable computational resources, suffer from unpredictable network latency. In addition, the mobility of users can lead to frequent connectivity disruptions, which make continuous access to cloud-based services unreliable.

To address these challenges, researchers have proposed a collaborative edge-device architecture that strategically deploys a small language model (SLM) on the device while offloading the large language model (LLM) to a base station (BS) or edge server~\cite{ding2024hybrid, hao2024hybrid, shao2025ai}. In~\cite{ding2024hybrid}, a router trained to predict query difficulty and desired quality level enables cost-efficient assignment of queries to either the small or large model. In~\cite{hao2024hybrid}, a cost-aware draft-verification approach was employed. By tuning a predefined threshold $p_t$ for the generated token probability, a controllable performance-cost trade-off was achieved.

However, these studies improve efficiency with a compromise of LLM inference accuracy. Therefore, \textbf{speculative decoding (SD)} was taken into account, where a small ``draft'' model generates $\gamma$ candidate tokens autoregressive, and then a big ``target'' model verifies these draft tokens in parallel~\cite{leviathan2023fast, chen2023accelerating}. In this way, the inefficiency of autoregressive token generation was mitigated without sacrificing the quality of inference. Furthermore, a \textbf{distributed speculative decoding (DSD)} architecture was first introduced in~\cite{zhao2024edge} with the draft model for token generation on the device and the target model for verification at the edge or base station (BS). The author tries to optimize the number of tokens generated by SLM to minimize delay and power consumption, taking the uplink and downlink transmission into consideration. 

Nevertheless, this hybrid deployment approach is constrained by communication bottlenecks: for each token, the device must transmit a full vocabulary distribution to the BS/edge server for LLM verification, resulting in a communication payload linearly dependent on vocabulary size. In~\cite{oh2024uncertainty}, the author proposed skipping uplink transmissions and LLM inference on tokens likely to be accepted. This improves token throughput but still at the expense of inference accuracy.

Building on previous works, we offer our solution: \textbf{a distributed split speculative decoding (DSSD)} framework. Specifically, SLM and LLM are still deployed on device and at edge, separately. But the SD verification phase was further split and distributed across the device and edge. By adopting this method, the uplink transmission of $\gamma$ vocabulary distributions from SLM is replaced by the downlink transmission of a single vocabulary distribution from LLM, which significantly reduces the communication time and payload.

The rest of the paper is organized as follows: Section~\ref{sec:System model} introduces the system model and existing DSD system. Section~\ref{sec:DSSD} provides our DSSD solutions. Section~\ref{sec:experiment} presents our experimental results and analysis. Conclusions are summarized in Section~\ref{sec:conclusion}. Appendix~\ref{sec:algorithm} provides the corresponding algorithms.

\section{System Model}\label{sec:System model}
% 这里我感觉还要描述得更清楚一点
We consider a collaborative inference architecture between a device and a BS. To achieve efficient collaborative inference, we deploy a lightweight small language model $M_q$ (SLM, denoted as $M_q$) on the device. Meanwhile, a computationally intensive large language model (LLM, denoted as $M_p$) is deployed in the BS ~\cite{zhao2024edge, oh2024uncertainty}.  We assume a vocabulary $\mathcal{V}$ is shared between SLM and LLM, where $\mathcal{V}$ represents the full set of possible tokens.

\subsection{Distributed Speculative Decoding}
\label{sec:DSD}
\begin{figure}[htbp]
\begin{center}
\centerline{\includegraphics[width=\columnwidth]{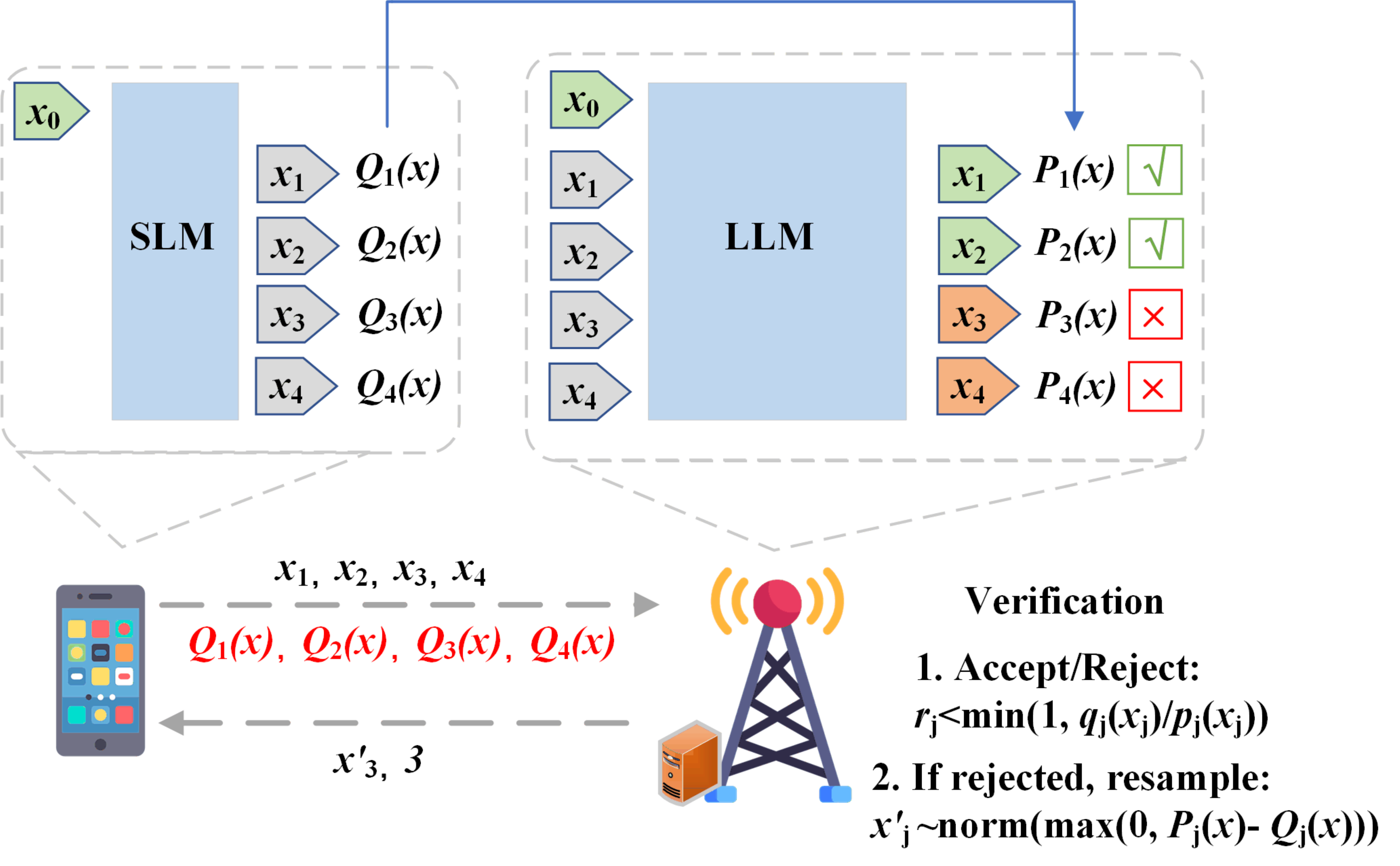}}
\caption{Distributed Speculative Decoding}
\label{fig:DSD}
\end{center}
\end{figure}

% 这里其实没有介绍清楚sp 的流程，要在jounal 中补充
Fig.~\ref{fig:DSD} illustrates the distributed speculative decoding (DSD) framework: SLM (on device) generates $\gamma$ draft tokens in an autoregressive manner. And LLM verifies them in parallel, accepting valid tokens and resampling new ones for rejected tokens. A detailed description is presented in Algorithm~\ref{alg:DSD} in the Appendix:

\textbf{Draft Process (on Deivce)}: 
In one single round, SLM generates $\gamma$ tokens based on the \textit{prefix}. To be more specific, for the $i$-th token, SLM first gets a vocabulary probability distribution, denoted as $Q_i(x)$ and then samples $x_i$ according to $Q_i(x)$, i.e., $x_i \sim Q_i(x)$.\footnote{$Q_i(x)$ or $P_i(x)$ is a vector with the same dimension as the vocabulary, i.e., $|\mathcal{V}|$.}

\textbf{Uplink Transmission}: 
The device sends the indices of the draft tokens with respect to the vocabulary $\mathcal{V}$ and vocabulary probability distributions to the BS.\footnote{Instead of transmitting full token strings, the device sends only the indices of draft tokens from the vocabulary, reducing communication overhead. For the sake of better illustration, however, tokens are consistently used in Fig.~\ref{fig:DSD} and Fig.~\ref{fig:DSSD}.}

\textbf{Verification Process (at the BS)}: LLM first gets the $\gamma+1$ distributions according to the prefix and received tokens: $P_j(x), j=1,\cdots,\gamma+1$. Then it verifies these received tokens: \textbf{\textit{a).~Accept/Reject:}} Let $q_j(x_j)$ and $p_j(x_j)$ denote the probability values of token $x_j$ in $Q_j(x)$ and $P_j(x)$. If $p_j(x_j)<q_j(x_i)$, $x_j$ is accepted as the $j$-th token. Otherwise, it is rejected with probability of $1-q_j(x_j)/p_j(x_j)$.\footnote{This is equivalent to the part of ``Accept'' in Algorithm.~\ref{alg:DSD}, i.e., $r_j< \min \left\{1, \frac{q_j(x_j)}{p_j(x_j)} \right\}$. The two different representations corresponds to \cite{leviathan2023fast} and \cite{chen2023accelerating}, respectively} \textbf{\textit{b).~Resample:}} Once rejected, it resamples a new token $x'_j \sim \mathrm{norm} \left(\max \left(0, P_j(x)-Q_j(x)\right)\right)$. If all $\gamma$ received tokens are accepted, it samples the $(\gamma+1)$-\textit{th} token $x_{\gamma+1} \sim P_{\gamma+1}(x)$.

\textbf{Downlink Transmission}: BS sends the results $x_j$ and $j$ back to the device, where $j$ denotes the position of the resampled token in the sequence if there is a rejection or equals $\gamma+1$ if all draft tokens are accepted.

\subsection{Wireless Communication}
To model the end-to-end communication delay, we take into account both the \textbf{transmission time (TT)} and the \textbf{non-transmission time (NTT)}, the latter comprising processing, propagation, and queuing times. Specifically, the transmission time consists of the uplink transmission time and downlink transmission time:
\begin{align}
    T_{up} &= \frac{D_{up}} {R_{up}}, \\
    T_{down} &= \frac{D_{down}}{R_{down}}.
\end{align}
where $D_{up}$ and $D_{down}$ are the amount of data transmitted in uplink and downlink, and are $R_{up}$ and $R_{down}$ are the transmission rate.

The round-trip time encompasses the cumulative delays of forward/return propagation, receiver processing, and queuing, which can be treated as a constant.

% The propagation time, on the other hand, accounts for the physical delay as the signal travels across the medium. It is determined by:
% \begin{equation}
%     T_{prop} = \frac{L}{v}
% \end{equation}
% where $L$ is the distance between device and BS, and 
% $v$ represents the propagation speed of the signal in the medium, typically close to the speed of light in air.

Thus, the communication time is
\begin{equation}
    T_{comm} = T_{up} + T_{down} + T_{NTT}
\end{equation}

Given that the index size is insignificant relative to the vocabulary distribution, our analysis considers only the uplink transmission latency associated with the vocabulary distribution. Hence, we have
\begin{equation}
    D_{up} = \gamma \cdot |\mathcal{V}| \cdot b_{prob}
\end{equation}
where $|\cdot|$ denotes the cardinality, $b_{prob}$ represents the bit-width of each probability value, e.g. $b_{prob}=32$ bits for full precision or 16 bits for half precision. 

The communication time becomes
\begin{equation}
    T_{comm} = \gamma \cdot \frac{|\mathcal{V}|b_{prob}}{R_{up}} + T_{NTT}
\end{equation}

\begin{table*}[t]
  \centering
  % 统一缩放到 0.8 倍宽度
  % \resizebox{1.1\linewidth}{!}{
    % 列间距略小，行高稍紧凑
    % \setlength{\tabcolsep}{3pt}
    % \renewcommand{\arraystretch}{1.2}
    % \scriptsize
    \begin{tabular}{c *{6}{c}}
      \toprule
      \diagbox{$\gamma$}{$\alpha$}
      % \(\gamma \,\backslash\, \alpha\) 
   & 0.5    & 0.6              & 0.7               & 0.8               & 0.9               & 0.99              \\
      \midrule
      2 & 0.75 (26.00\,ms) & 0.64 (25.12\,ms) & 0.51 (24.08\,ms) & 0.36 (22.88\,ms) & 0.19 (21.52\,ms) & 0.02 (20.16\,ms) \\
      4 & 0.94 (27.50\,ms) & 0.87 (26.96\,ms) & 0.76 (26.08\,ms) & 0.59 (24.72\,ms) & 0.34 (22.75\,ms) & 0.04 (20.32\,ms) \\
      6 & 0.98 (27.88\,ms) & 0.95 (27.63\,ms) & 0.88 (27.06\,ms) & 0.74 (25.90\,ms) & 0.47 (23.75\,ms) & 0.06 (20.47\,ms) \\
      8 & 1.00 (27.97\,ms) & 0.98 (27.87\,ms) & 0.94 (27.54\,ms) & 0.83 (26.66\,ms) & 0.57 (24.56\,ms) & 0.08 (20.62\,ms) \\
      \bottomrule
    \end{tabular}%
  % }
  \caption{The value of $1-\alpha^{\gamma}$ and expected communication time $T_{\mathrm{comm}}$.}
  \label{tab:Exoecred comm_time}
\end{table*}

\subsection{Wall-clock Time}
Inference latency comprises three parts: on-device SLM drafting time, edge-side LLM verification time, and device–edge communication time.

In a single run of draft-verify process, the inference latency is:
\begin{equation}
\label{eq:T_inf}
    T_{inf} = \gamma  \cdot T_{SLM} + T_{comm} + T_{LLM}
\end{equation}

where $T_q$ and $T_p$ denote the time for a single run of $M_p$ and $M_q$ respectively. With \eqref{eq:T_inf}, we have
\begin{equation}
\label{eq:T_inf2}
    T_{inf} =  \gamma  \cdot T_{SLM}  + T_{LLM} + \gamma  \cdot  \frac{|\mathcal{V}|b_{prob}}{R_{up}} + T_{NTT}
\end{equation}

% % Let
% \begin{align}
%     & b = T_{comm}/T_{SLM}; \\
%     & c = T_{SLM}/T_{LLM},
% \end{align}

% % We have

% \begin{equation}
%     T_{inf} = (c\gamma + b + 1) T_{LLM}.
% \end{equation}

% \begin{table*}[t]
% \centering
% \caption{Homogeneous experiment.  $b = T_{\mathrm{comm}}/T_{\mathrm{sLM}}$.}
% \label{tab:trend_table}
% \vskip 0.15in
% \begin{small}
% \begin{sc}
% \begin{tabular}{ccc|cc|cc|cc}
% \toprule
% \multirow{2.5}{*}{RTT (ms)} &
% \multirow{2.5}{*}{BW (Mbps)} &
% \multirow{2.5}{*}{$\gamma$} &
% \multicolumn{2}{c|}{Payload (KB)} &
% \multicolumn{2}{c|}{\shortstack{OPT-125M$\rightarrow$6.7B\\($\alpha=0.61\pm0.05$)}} &
% \multicolumn{2}{c}{\shortstack{OPT-125M$\rightarrow$13B\\($\alpha=0.53\pm0.05$)}} \\
% \cmidrule(lr){4-5}\cmidrule(lr){6-7}\cmidrule(lr){8-9}
%  & & & bit & dup\_B & $b$ & Speed & $b$ & Speed \\
% \midrule
%  0  & 100 & 8 & 61269 & 14.6$\downarrow$ & 0.00 & 2.28 & 0.00 & 1.88 \\
% 20  & 100 & 8 & _     & _               & 0.30 & 2.31 & 0.08 & 1.81 \\
% 50  &  10 & 8 & _     & _               & 0.70 & 2.19 & 0.23 & 1.73 \\
% \midrule
% 20  &  50 & 8 & 47130 & 100$\downarrow$ & 0.00 & 2.22 & 0.08 & 1.81 \\
% 50  &  50 & 8 & _     & _               & 0.00 & 1.97 & 0.22 & 1.74 \\
% 50  &  10 & 6 & _     & _               & 0.70 & 1.57 & 0.26 & 1.91 \\
% \bottomrule
% \end{tabular}
% \end{sc}
% \end{small}
% \vskip -0.1in
% \end{table*}

\section{Distributed Split Speculative Decoding}
\label{sec:DSSD}

\begin{figure}[htbp]
\begin{center}
\centerline{\includegraphics[width=0.8\columnwidth]{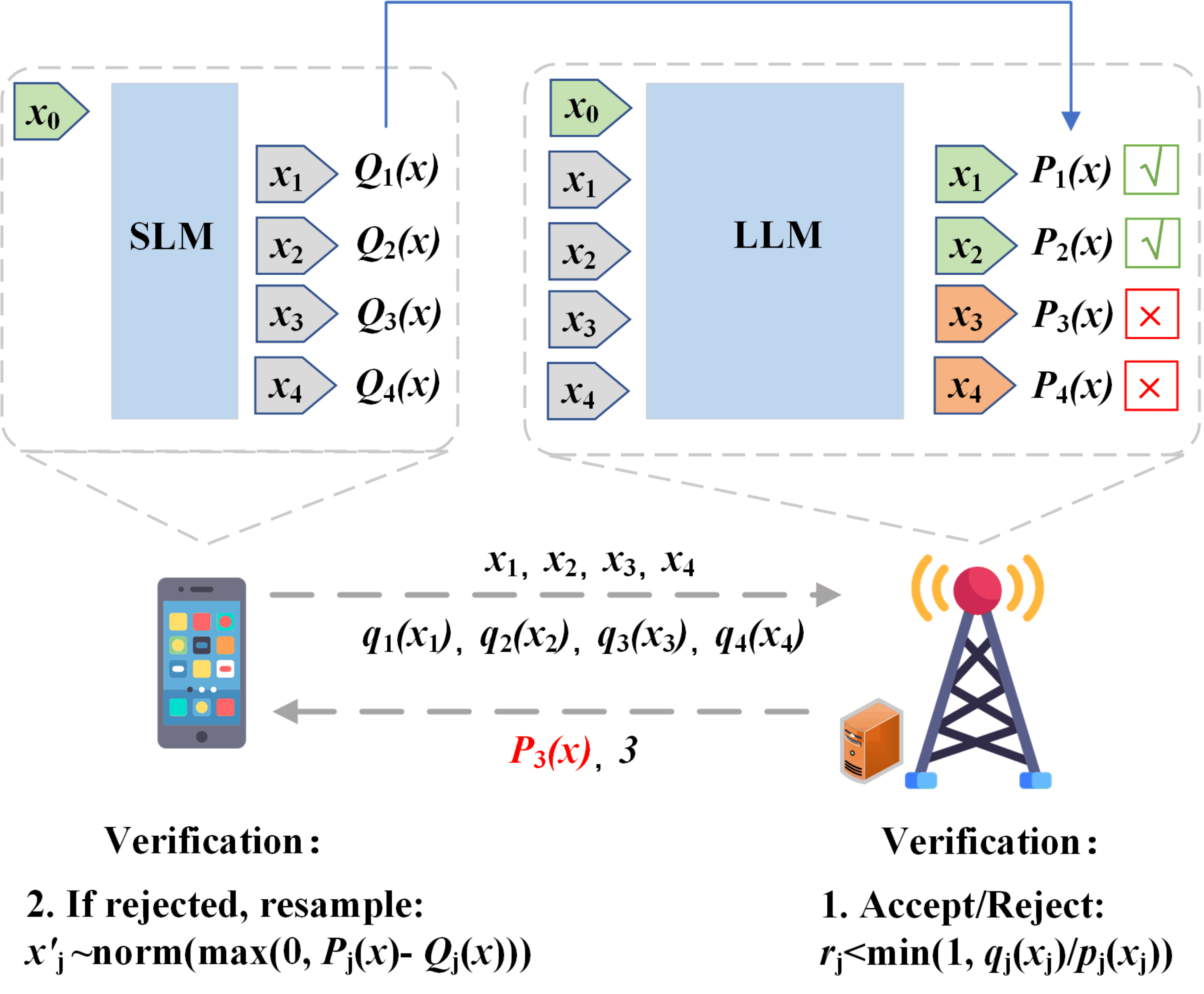}}
\caption{Distributed Split Speculative Decoding.}
\label{fig:DSSD}
\end{center}
\end{figure}
As shown in Fig.~\ref{fig:DSSD}, the distributed split speculative decoding (DSSD) employs the same framework as the DSD in Section.~\ref{sec:DSD}. The only distinction lies in the verification phase, which is split across the device and the edge. Refer to Algorithm~\ref{alg:DSSD} for details:

\textbf{Draft Process}: 
It is exactly the same as DSD in Section.~\ref{sec:DSD}.

\textbf{Uplink Transmission}: 
The device sends the indices and the probability values of the draft tokens with respect to the vocabulary $\mathcal{V}$ to the BS.

\textbf{Verification Process (at the BS)}: 
Only the process of \textbf{\textit{Accept/Reject}} is still handled by the LLM at the BS. And if all $\gamma$ received tokens are accepted, it samples the $\gamma+1$-th token $x_{\gamma+1} \sim P_{\gamma+1}(x)$.

\textbf{Downlink Transmission}: If there is a rejection, BS sends the vocabulary distribution $P_j(x)$ and $j$ back to the device, where $j$ denotes the position of the resampled token in the sequence. Otherwise (all draft tokens are accepted), it sends the $(\gamma+1)$-\textit{th} token along with the index $\gamma+1$.

\textbf{Verfication Process (on Device)}: the process of \textbf{\textit{Resample}} is done on device if there is a rejection, i.e., the received $j<\gamma+1$.

% In the previous distributed speculative decoding work, the SLM deployed on device generates $\gamma$ tokens per round and sends the draft tokens to the edge, along with $\gamma$ vocabulary distributions. The LLM deployed at edge verify these tokens in parallel: First, it decides whether accept or reject the token; Second, it resamples the token if rejected. In our Distributed Split Speculative Decoding (DSSD) framework, the verification was further split and distributed across the device and edge. Specifically, the \textit{accept/reject} phase is still on LLM but the \textit{resample} phase is done at SLM. Details are given in algorithms~1 and 2 at the appendix.

% To better illustrate our approach, a toy example is given in Fig.~\ref{fig:DSSD}. With our DSSD approach, only the vocabulary distribution of $P_3(x)$ is sent to the SLM via the downlink transmission. While for DSD, $4$ vocabulary distributions $Q_1(x), Q_2(x), Q_3(x), Q_4(x)$ are required to sent to the LLM via uplink transmission\footnote{The communication payload is primarily dominated by the vocabulary distribution due to its large size}.

\subsection{Wall-clock Time}
In DSSD, data transmission is primarily dominated by downlink traffic. The downlink transmission of vocabulary distribution occurs only when a rejection happens. Accordingly, the downlink transmission time can be expressed as
\begin{equation}
    T_{down} = a \cdot \frac{|\mathcal{V}|b_{prob}}{R_{down}}
\end{equation}
where $a = 1$ if there is a rejection, otherwise $a = 0$.

As the probability of all draft tokens being accepted is $\alpha^{\gamma}$, the average communication time is
\begin{equation}
    T_{comm} = (1-\alpha^{\gamma}) \cdot \frac{|\mathcal{V}|b_{prob}}{R_{down}} +T_{NTT}
\end{equation}

Hence, we have
\begin{corollary}
The communication time is bounded by
\begin{equation}
    T_{NTT} \leq T_{comm} \leq T_{down} + T_{NTT}
\end{equation}
\end{corollary}

As $\gamma \to \infty$, the probability of at least one rejection tends to 1.
For sufficiently large $\gamma$, the communication time could be approximated as 
\begin{equation}
    T_{comm} \approx T_{down} +T_{NTT}
\end{equation}

On the other hand, as $\alpha \to 1$, all tokens will be accepted. And.
\begin{equation}
    T_{comm} \approx T_{NTT}
\end{equation}

Table.~\ref{tab:Exoecred comm_time} shows the value of  $1-\alpha^{\gamma}$ and communication time under different $\alpha$ and $\gamma$.

The inference latency is further expressed as:
\begin{equation}
    T_{inf} \!=\! \gamma  \cdot T_{SLM} + T_{LLM} + (1-\alpha^{\gamma}) \cdot \frac{|\mathcal{V}|b_{prob}}{R_{down}} +T_{NTT}
\end{equation}

% 仿真用表，直接生成的
% Similarly to DSD, 
\section{Experiment}\label{sec:experiment}

To assess the proposed “Distributed Split Speculative Decoding” framework, we designed and employed an end-to-end experimental setup that simulates the computational heterogeneity of devices and the base station, as well as the network conditions between them. Our main evaluation metric is the \textbf{\textit{Speedup ratio}}, which quantifies how much the throughput improves when using the existing DSD or proposed DSSD method compared to conventional LLM inference. That is
\begin{equation}
    Speedup = \frac{TP_{DSD}}{TP_{LLM}}
\end{equation}
or
\begin{equation}
    Speedup = \frac{TP_{DSSD}}{TP_{LLM}}
\end{equation}
where $TP_{DSD}$, $TP_{DSSD}$ and $TP_{LLM}$ are the throughputs of DSD, DSSD and LLM, respectively.

In this setup, two types of GPUs are configured to emulate the computational heterogeneity. On the device side, a lightweight draft model (OPT-$125$M) generates speculative draft tokens, while the base station employs a larger target model (OPT-$13$B or OPT-$6.7$B) to verify and correct any token prediction errors.
Controllable latency and throughput limits were introduced into the communication between the device and BS to simulate real link conditions. The round-trip latency was configured as one of the values in $\{0, 20, 50\}$~ms, and the bandwidth capped at $100$~Mbps~\cite{affandi2024throughput, ateya2018study}.

Each experiment begins with a common 128-token narrative prompt and generates an additional 128 tokens using the top-k sampling with a temperature of 1.0 and $k=10$, in order to maintain full distribution fidelity.

% Our main goals have four aspects:
% 6.7b 模型的平均轮次是30轮，13B模型的平均轮次
\subsection{Homogeneous device}
In a homogeneous experiment, two H800 chips are deployed. 
One runs a small OPT-125M draft model on the device side\footnote{Assume the device operates under limited memory and computational capacity.}, while the other runs a large OPT-13B or OPT-6.7B verification model on the base station side. This setup isolates the effect of heterogeneous devices on speculative decoding throughput while still accounting for the latency introduced by the wireless environment. Table~2 presents the expected acceleration speed, predicted by our analytical model, and the empirical acceleration speed measured end-to-end on homogeneous hardware. Table~\ref{tab:trend_table} presents the expected acceleration speed, predicted by our analytical model, and the empirical acceleration speed measured end-to-end on homogeneous hardware.
\begin{table*}[h]
\centering
\vskip 0.15in
\begin{small}
\begin{sc}
\begin{tabular}{ccc|cc|cc|cc}
\toprule
\multirow{2.5}{*}{NTT (ms)} &
\multirow{2.5}{*}{BW (Mbps)} &
\multirow{2.5}{*}{$\gamma$} &
\multicolumn{2}{c|}{\shortstack{OPT-125M$\rightarrow$6.7B \\ DSD($\alpha=0.61\pm0.05$) }} &
\multicolumn{2}{c|}{\shortstack{OPT-125M$\rightarrow$6.7B \\ DSSD($\alpha=0.61\pm0.05$)}} &
\multicolumn{2}{c}{\shortstack{OPT-125M$\rightarrow$13B \\ DSSD($\alpha=0.53\pm0.05$)}} \\
\cmidrule(lr){4-5}\cmidrule(lr){6-7}\cmidrule(lr){8-9}
 & & & $T_{com}$(s) & Speedup & $T_{com}$(s) & Speedup & $T_{com}$(s) & Speedup \\
\midrule
 0  & 100 & 8 & 3.91 & 1.51  & 0.01 & 2.28 & 1.40 & 1.67 \\
20  & 100 & 8 & 4.60 & 1.41  & 0.62 & 2.31 & 9.21 & 1.72 \\
50  &  10 & 8 & 41.44& 0.43  & 1.55 & 2.19 & 2.30 & 1.62 \\
\midrule
20  &  50 & 8 & 8.59  & 1.24  & 0.62 & 2.22 & 0.92 & 1.77 \\
50  &  50 & 8 & 9.52  & 1.08  & 1.55 & 1.97 & 2.30 & 1.70 \\
50  &  10 & 6 & 8.12  & 1.16  & 0.40 & 1.57 & 2.01 & 1.97 \\
\bottomrule
\end{tabular}
\end{sc}
\end{small}
% \vskip -0.1in
\caption{Homogeneous experiment.}
\label{tab:trend_table}
\end{table*}

% DSD -> DSSD 的转变
As shown in the table.\ref{tab:trend_table}, the communication time of the DSSD method is much less than that of DSD. 
The amount of data uploaded by the DSD method is approximately 61,269 bytes per round, in contrast to the DSSD method, which uploads less than 50 bytes each time.
This is because the DSSD method converts the draft model’s ``overloaded'' uplink transmission into the target model's ``minimal'' downlink transmission, which significantly reduces the communication load. 
As a result, the DSSD method achieves a SPEED much greater than DSD, thus delivering higher overall inference performance. Specifically, the speed-up of the DSSD method is between $1.5\times$ and $2.4\times$, while the speed-up of DSD only remains at $1\times$ or even lower. This is the direct result after fully considering the communication cost.

% 通信量变重的转变，关于b 值的改变带来的影响
When the draft length is fixed at $\gamma = 8$ for the OPT-125M$\rightarrow$6.7B pair, the Speedup ratio shows significant variation for DSD, in contrast to the relatively consistent performance of DSSD. While for DSSD with different targe models, i.e., OPT-125M$\rightarrow$6.7B and OPT-125M$\rightarrow$13B, the Speedup ratio declines with a lower acceptance rate, primarily due to increased misalignment resulting from the larger model size.

\subsection{Heterogeneous device}
In heterogeneous experiments, we deployed a small OPT-125M draft model on an NVIDIA A6000 GPU. Meanwhile, the large OPT-13B or OPT-6.7B verification model ran on an NVIDIA H800 GPU.
As DSSD has demonstrated superior performance over DSD in the preceding experiments, our comparisons are restricted to variations in the target LLM’s model size.

To better analyze the experimental results, we first introduce and compute the ratio $c = T_ {SLM} / T_ {LLM}$ as defined in \cite{leviathan2023fast}. For the 125M$\rightarrow$6.7B configuration, $c$ is approximately $0.1$, while for 125M$\rightarrow$13B, $c$ is about $0.05$. As a result, the SLM's multiple executions are masked by a single LLM decoding. Fixing the draft length at $\gamma = 6$ and varying the link conditions shows that an ideal $0\,\text{ms}/100\,\text{Mbps}$ channel yields a modest $1.02\times$ speed-up for the OPT-6.7B target model, and a larger $1.48\times$ for OPT-13B. With a $20$ms NTT at the same bandwidth, the advantage shifts to the larger model, whose throughput peaks at $1.75$, whereas the smaller model’s throughput drops to $0.92$. When bandwidth is throttled to $50\,\text{Mbps}$ under the same latency, the trend reverses: OPT-6.7 B rebounds to $1.38\times$, whereas OPT-13 B falls to $1.24\times$, underscoring the latter’s bandwidth sensitivity.

% To introduce the important calculation than $c = T_ {SLM} / T_ {LLM} $, the heterogeneous experiment uniformly employed the DSSD method, and did not explicitly display the changes of $b$ (because the DSSD method had already demonstrated its superior performance in terms of bandwidth in the previous experiment). 
% Table \ref{tab:hetero_gamma6}. Baseline autoregressive throughput and end-to-end speculative decoding throughput measured under three typical network conditions That is, the ideal (0 ms/100 Mbps), medium (20 ms/50 Mbps) and high overhead (50 ms/10 Mbps) batch processing size $\gamma\in\{2,4,6,8\}$. The range of the calculation ratio $c$ decreased from approximately 0.566 ($\gamma=2$) to 0.288 ($\gamma=8$) for the 6.7B pipeline, and from 0.470 to 0.215 (13B pipeline), reflecting the increased draft cost on the weaker GPU.
% 检查这里的C是否正确，好像有点不对

% In the heterogeneous setting where an NVIDIA A6000 generates drafts and an H800 finalizes them, the compute-time ratio is approximately $c = 0.1$, so network latency and bandwidth dominate once SLM execution is hidden behind LLM decoding.  Fixing the draft length at $\gamma = 6$ and sweeping the link shows that an ideal $0\,\text{ms}/100\,\text{Mbps}$ channel yields a modest $1.02\times$ speed-up for the OPT-6.7 B target and a larger $1.48\times$ for OPT-13 B.  
% Introducing a $20\,\text{ms}$ RTT at the same bandwidth shifts the advantage to the larger model, whose throughput peaks at $1.75\times$ while the smaller model drops to $0.92\times$.  

Holding the link at $20\,\text{ms}/100\,\text{Mbps}$ while varying the draft length reveals an interior optimum. A short draft ($\gamma = 4$) does not sufficiently amortize the communication delay, limiting speedups to $0.82\times$ for 6.7B and $1.37\times$ for 13B. Extending the draft to $\gamma = 6$ balances synchronization costs and speculative overhead, achieving global maxima of $0.92\times$ and $1.75\times$.  
Increasing the draft further to $\gamma = 8$ overloads the speculative path, reducing speedups to $0.87\times$ and $1.14\times$. Overall, these results demonstrate that even on an asymmetric A6000–H800 pair, speculative decoding remains effective: with proper tuning of the draft length and moderate network resources, throughput gains of up to $1.8\times$ for OPT-13B and $1.4\times$ for OPT-6.7B can be sustained under realistic communication conditions.

% It is worth noting that the performance gap between heterogeneous devices remains within $10\times$. This is a very important condition because if the performance gap between two devices is too large, the significance of sampling acceleration is lost. And determining the boundaries for when to conduct speculative sampling and when to perform task offloading will be the focus of our next work.

% In the top module ($\gamma=6$), the DSP throughput of OPT-125 M→6.7 B decreased from 1.98 tok/s in the zero-latency case to 1.71 tok/s in the 50 ms/10 Mbps case, while the baseline LLM remained at ~ 7 tok/s. Increasing $\gamma$to 4 (the middle block) allows the 6.7B path to recover the amortized fixed communication delay to 2.32 tok/s at 20 ms/50 Mbps, and even higher at 0 ms/100 Mbps. For the OPT-125 M→ 1.3B pairing, under medium and low-overhead links, the lower computational ratio $c$and the consistently high acceptance rate ($\alpha\approx1.0$and $\gamma=2$) translate into a DSP throughput of more than 4 tok/s.

\begin{table}[t]
\centering
\vskip 0.15in
\begin{small}
\begin{sc}
\resizebox{\linewidth}{!}{
    \begin{tabular}{ccc|c|c|c}
    \toprule
    \multirow{2.5}{*}{NTT (ms)}&\multirow{2.5}{*}{BW (Mbps)} & \multirow{2.5}{*}{$\gamma$} & \multicolumn{1}{c}{6.7B(DSD)} &\multicolumn{1}{c}{125M$\rightarrow$6.7B} & \multicolumn{1}{c}{125M$\rightarrow$13B} \\
    \cmidrule(lr){4-6}
             &       &      & Speedup & Speedup         & Speedup\\
    \midrule
    0        & 100    &  8  & 0.88  & 0.98           & 1.17\\
    20       & 100    &  8  & 0.85  & 0.92           & 1.13\\
    20       & 50     &  8  & 0.72  & 0.92               & 1.12\\
    \midrule
    20       & 100     &  4  & 1.10 & 1.11           & 1.44\\
    20       & 100     &  6  & 1.12 & 1.14           & 1.58\\
    20       & 100     &  8  & 0.93 & 1.04           & 1.30\\
    \bottomrule
    \end{tabular}
}
\end{sc}
\end{small}
\caption{Heterogeneous experiment.}
\label{tab:hetero_gamma6}
\end{table}

\section{Conclusion}
\label{sec:conclusion}
This work presents an early but meaningful exploration of distributed LLM inference using speculative decoding. Under the current settings, communication time is the main bottleneck for throughput. To reduce communication overhead while maintaining inference performance, the DDSD solution was proposed: the verification process is divided into two stages—``Accept/Reject'' and ``Resample''—which are executed at the BS and on the device, respectively. In this way, the uplink transmission of $\gamma$ probability distribution was reduced to $\gamma$ token index and probability values, and the downlink transmission of one probability distribution at most. Experimental evaluations provide empirical support for these design choices
and offer insights into throughput behavior under different draft length and communication conditions.

% It introduces a new perspective, that is, to infer how decoding can be scaled across heterogeneous systems, significantly reducing communication overhead while retaining most of the acceleration advantages. 
% In particular, the core idea of transmitting only $\gamma$ token probability pairs (rather than a full probability distribution) provides a simple yet effective reduction in uplink traffic, achieving several orders of magnitude in efficiency improvement. The accompanying prototypes and experimental evaluations provide empirical support for these design choices and offer insights into throughput behavior under different $\gamma$, bandwidth and latency conditions.

% in the current setup, communication modeling is still relatively simplified. The network latency is approximately simulated using functions, and real dynamics such as jitter, packet loss or protocol overhead have not been captured yet. Therefore, although the experimental results are encouraging, they represent an idealized baseline. Future work should focus on incorporating a more realistic communication layer and exploring deployment scenarios under real network conditions. Through more refined communication models and expanded test environments, DSSD has great potential as a practical method to accelerate the inference of large language models across edge and cloud systems.

\section*{Acknowledgement}
This work was supported by the National Key Research and Development Program of China under Grant No.2024YFE0200800. The methods presented in this work are covered by a filed Patent with Application No.CN2025108856278.

% \section*{Acknowledgments}
% Fundings:.

\bibliography{main_ICML}

\begin{thebibliography}{10}
\providecommand{\natexlab}[1]{#1}
\providecommand{\url}[1]{\texttt{#1}}
\expandafter\ifx\csname urlstyle\endcsname\relax
  \providecommand{\doi}[1]{doi: #1}\else
  \providecommand{\doi}{doi: \begingroup \urlstyle{rm}\Url}\fi

\bibitem[Affandi et~al.(2024)Affandi, Riyadi, and Prakoso]{affandi2024throughput}
Affandi, M.~A., Riyadi, M.~A., and Prakoso, T.
\newblock Throughput and coverage evaluation on the use of existing cellular towers for 5g network in surakarta city.
\newblock \emph{Jurnal Ilmiah Teknik Elektro Komputer dan Informatika (JITEKI)}, 10\penalty0 (1):\penalty0 54--72, 2024.

\bibitem[Ateya et~al.(2018)Ateya, Muthanna, Makolkina, and Koucheryavy]{ateya2018study}
Ateya, A.~A., Muthanna, A., Makolkina, M., and Koucheryavy, A.
\newblock Study of 5g services standardization: Specifications and requirements.
\newblock In \emph{2018 10th international congress on ultra modern telecommunications and control systems and workshops (ICUMT)}, pp.\  1--6. IEEE, 2018.

\bibitem[Chen et~al.(2023)Chen, Borgeaud, Irving, Lespiau, Sifre, and Jumper]{chen2023accelerating}
Chen, C., Borgeaud, S., Irving, G., Lespiau, J.-B., Sifre, L., and Jumper, J.
\newblock Accelerating large language model decoding with speculative sampling.
\newblock \emph{arXiv preprint arXiv:2302.01318}, 2023.

\bibitem[Chen et~al.(2024)Chen, Yang, Chong, and Quek]{chen2024personalizing}
Chen, Z., Yang, H.~H., Chong, K. F.~E., and Quek, T. Q.~S.
\newblock Personalizing semantic communication: A foundation model approach.
\newblock In \emph{2024 {IEEE} Int. Workshop Signal Process. Adv. Wireless Commun. (SPAWC)}, pp.\  846--850. IEEE, 2024.

\bibitem[Ding et~al.(2024)Ding, Mallick, Wang, Sim, Mukherjee, Ruhle, Lakshmanan, and Awadallah]{ding2024hybrid}
Ding, D., Mallick, A., Wang, C., Sim, R., Mukherjee, S., Ruhle, V., Lakshmanan, L.~V., and Awadallah, A.~H.
\newblock Hybrid llm: Cost-efficient and quality-aware query routing.
\newblock \emph{arXiv preprint arXiv:2404.14618}, 2024.

\bibitem[Hao et~al.(2024)Hao, Jiang, Jiang, Ren, and Cao]{hao2024hybrid}
Hao, Z., Jiang, H., Jiang, S., Ren, J., and Cao, T.
\newblock Hybrid slm and llm for edge-cloud collaborative inference.
\newblock In \emph{Proceedings of the Workshop on Edge and Mobile Foundation Models}, pp.\  36--41, 2024.

\bibitem[Leviathan et~al.(2023)Leviathan, Kalman, and Matias]{leviathan2023fast}
Leviathan, Y., Kalman, M., and Matias, Y.
\newblock Fast inference from transformers via speculative decoding.
\newblock In \emph{International Conference on Machine Learning}, pp.\  19274--19286. PMLR, 2023.

\bibitem[Oh et~al.(2024)Oh, Kim, Park, Ko, Quek, and Kim]{oh2024uncertainty}
Oh, S., Kim, J., Park, J., Ko, S.-W., Quek, T.~Q., and Kim, S.-L.
\newblock Uncertainty-aware hybrid inference with on-device small and remote large language models.
\newblock \emph{arXiv preprint arXiv:2412.12687}, 2024.

\bibitem[Shao \& Li(2025)Shao and Li]{shao2025ai}
Shao, J. and Li, X.
\newblock Ai flow at the network edge.
\newblock \emph{IEEE Network}, 2025.

\bibitem[Zhao et~al.(2024)Zhao, Jing, Lu, and Wen]{zhao2024edge}
Zhao, W., Jing, W., Lu, Z., and Wen, X.
\newblock Edge and terminal cooperation enabled {llm} deployment optimization in wireless network.
\newblock In \emph{2024 IEEE/CIC International Conference on Communications in China (ICCC Workshops)}, pp.\  220--225. IEEE, 2024.

\end{thebibliography}
\bibliographystyle{icml2025}

%%%%%%%%%%%%%%%%%%%%%%%%%%%%%%%%%%%%%%%%%%%%%%%%%%%%%%%%%%%%%%%%%%%%%%%%%%%%%%%
%%%%%%%%%%%%%%%%%%%%%%%%%%%%%%%%%%%%%%%%%%%%%%%%%%%%%%%%%%%%%%%%%%%%%%%%%%%%%%%
% APPENDIX
%%%%%%%%%%%%%%%%%%%%%%%%%%%%%%%%%%%%%%%%%%%%%%%%%%%%%%%%%%%%%%%%%%%%%%%%%%%%%%%
%%%%%%%%%%%%%%%%%%%%%%%%%%%%%%%%%%%%%%%%%%%%%%%%%%%%%%%%%%%%%%%%%%%%%%%%%%%%%%%
\newpage
\appendix
\onecolumn
\section{Algorithms}
\label{sec:algorithm}
Algorithm~\ref{alg:DSD} gives the full details of Distributed Speculative Decoding (DSD). And Algorithms~\ref{alg:DSSD} gives the full details of Distributed Split Speculative Decoding (DSSD).

\begin{algorithm}[ht]
\caption{Distributed Speculative Decoding (DSD)}
\label{alg:DSD}
\begin{algorithmic}

% Draft Process on device
\STATE \hspace{-1.2em} \textcolor{blue}{$\triangleright$ Draft Process}
\STATE \hspace{-1.2em} {\bfseries Device:} Initialize $\mathbf{y}=[]$
    \STATE {\bfseries Input:} prefix, SLM -- $M_q$, token length -- $\gamma$
% Autoregressive Generation
    \STATE \textcolor{blue}{$\triangleright$ Sample $\gamma$ tokens autoregressively}
    \FOR{$i=1$ {\bfseries to} $\gamma$}
    \STATE $Q_i(x)\leftarrow M_q(prefix+\mathbf{y})$;\\
    Sample $x_i \sim Q_i(x)$;\\
    $\mathbf{y} = \mathbf{y}+x_i$
    \ENDFOR
    \STATE {\bfseries Output:} $[x_1,\cdots,x_{\gamma}]$ and $[Q_1(x), \cdots, Q_{\gamma}(x)]$
\STATE \vspace{-2mm} \hspace{-1em}\hrulefill

% Uplink Transmission
\STATE \hspace{-1.2em} \textcolor{blue}{$\triangleright$ Uplink Transmission}
    \STATE Send tokens $[x_1,\cdots,x_{\gamma}]$ and {\color{red} probability distributions $[Q_1(x), \cdots, Q_{\gamma}(x)]$} to the BS via uplink transmission\footnotemark
\STATE \vspace{-2mm} \hspace{-1em}\hrulefill

% Verification at BS
\STATE \hspace{-1.2em} \textcolor{blue}{$\triangleright$ Verification Process}
\STATE \hspace{-1.2em} {\bfseries Edge:} Initialize $j=1$, $\mathrm{Flag} = 1$
    \STATE {\bfseries Input:} prefix, LLM -- $M_p$
% Parallel decoding
    \STATE \textcolor{blue}{$\triangleright$ Run $M_p$ in parallel}
    \STATE $P_1(x), \cdots, P_{\gamma+1}(x) \leftarrow$
    $M_q(prefix), \cdots, M_q(prefix + [x_1,\cdots, x_{\gamma}])$
    \WHILE{$j \leq \gamma$ \& $\mathrm{Flag}=1$}
        \STATE Sample $r_j \sim U[0,1]$ from a uniform distribution.
        \STATE \textcolor{red}{$\triangleright$ Accept}
        \IF{$r_j< \min \left\{1, \frac{q_j(x_j)}{p_j(x_j)} \right\}$}
        \STATE $x_j$ is accepted;\\ 
        $j=j+1$
        \STATE \hspace{-1.2em} \textcolor{red}{$\triangleright$ Reject and Resample}
        \ELSE
        \STATE $x'_j \sim \mathrm{norm} \left(\max \left(0, P_j(x)-Q_j(x)\right)\right)$;\\
        $x_j = x'_j$;\\
        $\mathrm{Flag}=0$
        \ENDIF
    \IF{$j=\gamma+1$}
        \STATE Sample $x_{\gamma+1} \sim P_{\gamma+1}(x)$
    \ENDIF
    \ENDWHILE
\STATE \vspace{-2mm} \hspace{-1em}\hrulefill

% Downlink Transmission
\STATE \hspace{-1.2em} \textcolor{blue}{$\triangleright$ Downlink Transmission}
    \STATE Send $x_j$ and $j$
\STATE \vspace{-2mm} \hspace{-1em}\hrulefill
\STATE \hspace{-1.2em} \textcolor{blue}{$\triangleright$ Reset}
\STATE Let $\mathrm{prefix} = \mathrm{prefix} + [x_1, \cdots, x_{j}]$ for Device and Edge
\end{algorithmic}
\end{algorithm}
\footnotetext{The device actually sends the indices of the draft tokens $[x_1,\cdots,x_{\gamma}]$ from the Vocabulary $\mathcal{V}$.}

\begin{algorithm}[ht]
\caption{Distributed Split Speculative Decoding (DSSD)}
\label{alg:DSSD}
\begin{algorithmic}

% Draft Process on device
\STATE \hspace{-1.2em} \textcolor{blue}{$\triangleright$ Draft Process}
\STATE \hspace{-1.2em} {\bfseries Device:} Initialize $\mathbf{y}=[]$
    \STATE {\bfseries Input:} prefix, SLM -- $M_q$, token length -- $\gamma$
% Autoregressive Generation
    \STATE \textcolor{blue}{$\triangleright$ Sample $\gamma$ tokens autoregressively}
    \FOR{$i=1$ {\bfseries to} $\gamma$}
    \STATE $Q_i(x)\leftarrow M_q(prefix+\mathbf{y})$;\\
    Sample $x_i \sim Q_i(x)$;\\ 
    $\mathbf{y} = \mathbf{y}+x_i$
    \ENDFOR
    \STATE {\bfseries Output:} $[x_1,\cdots,x_{\gamma}]$ and $[Q_1(x), \cdots, Q_{\gamma}(x)]$
\STATE \vspace{-2mm} \hspace{-1em}\hrulefill

% Uplink Transmission
\STATE \hspace{-1.2em} \textcolor{blue}{$\triangleright$ Uplink Transmission}
    \STATE Send tokens $[x_1,\cdots,x_{\gamma}]$ and {\color{red} probability values $[q_1(x), \cdots, q_{\gamma}(x)]$} to the BS via uplink transmission
\STATE \vspace{-2mm} \hspace{-1em}\hrulefill

% Verification at BS
\STATE \hspace{-1.2em} \textcolor{blue}{$\triangleright$ Verification Process}
\STATE \hspace{-1.2em} {\bfseries Edge:} Initialize $j=1$, $\mathrm{Flag} = 1$
    \STATE {\bfseries Input:} prefix, LLM -- $M_p$
% Parallel decoding
    \STATE \textcolor{blue}{$\triangleright$ Run $M_p$ in parallel}
    \STATE $P_1(x), \cdots, P_{\gamma+1}(x) \leftarrow$
    $M_q(prefix), \cdots, M_q(prefix + [x_1,\cdots, x_{\gamma}])$
    \WHILE{$j \leq \gamma$ \& $\mathrm{Flag}=1$}
        \STATE Sample $r_j \sim U[0,1]$ from a uniform distribution.
        \STATE \textcolor{red}{$\triangleright$ Accept}
        \IF{$r_j< \min \left\{1, \frac{q_j(x_j)}{p_j(x_j)} \right\}$}
        \STATE $x_j$ is accepted;\\
        $j=j+1$
        \STATE \hspace{-1.2em} \textcolor{red}{$\triangleright$ Reject}
        \ELSE
        \STATE $\mathrm{Flag}=0$
        \ENDIF
    \IF{$j=\gamma+1$}
        \STATE Sample $x_{\gamma+1} \sim P_{\gamma+1}(x)$
    \ENDIF
    \ENDWHILE

\STATE \vspace{-2mm} 
\hspace{-1em}\hrulefill

% Downlink Transmission
\STATE \hspace{-1.2em} \textcolor{blue}{$\triangleright$ Downlink Transmission}
    \STATE Send {\color{red}  $P_j(x)$} and $j$ if $\mathrm{Flag}=0$ or $x_j$ and $j$ if $\mathrm{Flag}=1$ 
\STATE \vspace{-1mm} \hspace{-1em}\hrulefill

\STATE \hspace{-1.2em} \textcolor{red}{$\triangleright$ Resample}
\STATE \hspace{-1.2em} {\bfseries Device:} 
    \IF{$\mathrm{Flag}=0$}
        \STATE $x'_j \sim \mathrm{norm} \left(\max \left(0, P_j(x)-Q_j(x)\right)\right)$;\\
        $x_j = x'_j$; 
    \ENDIF
\STATE \vspace{-2mm} 
\hspace{-1em}\hrulefill

\STATE \hspace{-1.2em} \textcolor{blue}{$\triangleright$ Reset}
\STATE Let $\mathrm{prefix} = \mathrm{prefix} + [x_1, \cdots, x_{j}]$ for Device and Edge.\\
\textbf{Note}: the device should still need to upload the resampled $x_j$ to edge if $\mathrm{Flag} = 0$. But this can be done in the next round of draft-verify process

\end{algorithmic}
\end{algorithm}

\end{document}